\documentclass[pdflatex,sn-mathphys-num]{sn-jnl}

\usepackage{amsmath,amssymb,amsfonts}
\usepackage{amsthm}
\usepackage{mathrsfs}

\usepackage[T1]{fontenc}
\usepackage[utf8]{inputenc}

\usepackage{booktabs}
\usepackage{multirow}
\usepackage{array}
\usepackage{makecell}
\usepackage{tabularx}

\usepackage{graphicx}
\usepackage{subcaption}
\usepackage[labelfont=bf]{caption}

\usepackage{xcolor}

\usepackage{textcomp}

\usepackage{hyperref}
\usepackage{url}

\usepackage[title]{appendix}

\usepackage{tikz}

\theoremstyle{thmstyleone}

\theoremstyle{thmstyletwo}

\theoremstyle{thmstylethree}

\begin{document}

\title[Article Title]{ProPACT: A Proactive AI-Driven Adaptive Collaborative Tutor for Pair Programming}


\author*[1]{\fnm{Anahita} \sur{Golrang}}\email{anahita.m.golrang@ntnu.no}

\author[1]{\fnm{Kshitij } \sur{Sharma}}

\author[1]{\fnm{Simon } \sur{Dehaen}}

\author[2]{\fnm{Olga } \sur{Viberg}}

\affil*[1]{\orgdiv{Department of Computer Science}, \orgname{Norwegian University of Science and Technology}, \orgaddress{ \city{Trondheim},  \country{Norway}}}

\affil[2]{\orgdiv{Department}, \orgname{The Royal Institute of Technology (KTH), School of Electrical Engineering and Computer Science}}



\abstract{Effective pair programming depends on coordination of attention, cognitive effort, and joint regulation over time, yet most adaptive learning systems remain individual-centric and reactive. This paper introduces ProPACT, a proactive AI-driven adaptive collaborative tutor that treats collaboration itself as the object of instruction. ProPACT constructs a multimodal dyadic learner model based on Joint Visual Attention (JVA), Joint Mental Effort (JME), and individual mental effort, and employs an XGBoost-based forecasting model to predict emerging suboptimal collaboration states up to 30 seconds in advance. These predictions drive a hierarchical adaptive policy that delivers minimally intrusive scaffolds while fading support during productive collaboration. A within-subject study with 26 pair-programming dyads shows that proactive feedback significantly improves debugging success, task efficiency, feedback uptake, and post-intervention gains in JVA and JME, demonstrating the potential of forecast-driven dyadic adaptivity for real-time collaborative learning regulation.}



\maketitle

\backmatter




\section{Introduction and Background}
\subsection{{Collaborative Programming as a Dyadic Regulation Challenge}}
Collaborative programming is a cognitively demanding learning activity in which success depends not only on individual programming ability, but on how effectively partners coordinate attention, balance cognitive effort, and regulate their shared problem-solving processes over time \cite{sharma2023happens}.
In pair programming–a common type of collaborative programming, learners must continuously align their understanding, negotiate roles, monitor progress, and repair breakdowns in coordination \cite{rodriguez2017exploring}. These processes are supported by socially shared regulation, co-regulation, and group metacognition, which play a central role in sustaining productive collaboration and learning \cite{jarvela2020bridging, jarvela2019capturing, liu2023collaborative}.
However, collaborative regulation is inherently dynamic, fragile, and difficult to sustain, particularly in technology-mediated environments \cite{bakhtiar2020dynamic}. Breakdowns in attention alignment, workload balance, or shared understanding can quickly degrade both collaboration quality and learning outcomes \cite{jarvela2019capturing}. Supporting these processes in real time remains a core challenge for Artificial Intelligence in Education (AIED) scholars and practitioners.
\subsection{{Attentional and Cognitive Mechanisms in Collaborative Learning}}
Effective collaboration relies on the coordination of both attention  and cognitive effort \cite{sharma2022brings}. One key mechanism is Joint Visual Attention (JVA), i.e., to the extent to which partners attend to the same task-relevant information at the same time \cite{ke2024using}. Sustained JVA supports grounding, shared situational awareness, and coordinated action, which are essential for collaborative problem solving \cite{sharma2023happens}.
A second mechanism concerns mental effort and cognitive load. At the individual level, mental effort reflects a learner’s cognitive workload and engagement, providing insight into potential overload or disengagement \cite{bueno2021effects,polat2025learning}. Yet in collaborative contexts, learning depends not only on individual effort, but on how partners’ cognitive effort aligns over time \cite{sharma2023happens}. This dyadic alignment can be captured through Joint Mental Effort (JME), which reflects the synchrony and balance of partners’ cognitive engagement \cite{sharma2022brings}. While individual mental effort reveals each learner’s cognitive state, JME provides a direct indicator of collaborative regulation and cognitive coordination—key drivers of effective joint problem solving \cite{sharma2021challenging, d2021shared}.
Together, JVA, individual mental effort, and JME offer theoretically grounded indicators of shared attention, cognitive balance, and collaborative regulation.
\subsection{{Multimodal Learning Analytics for Modeling Dyadic Learning Processes}}
Advances in Multimodal Learning Analytics (MLA) have expanded the ability to observe and model these otherwise invisible collaborative processes \cite{sharma2020multimodal,spikol2017current,sharma2019building}. Dual eye-tracking enables fine-grained measurement of attentional alignment and real-time estimation of JVA, while pupillometry provides continuous proxies for individual cognitive load and mental effort \cite{sharma2021challenging}. Combining these modalities enables the construction of dyadic learner models that represent both individual and shared cognitive dynamics. Prior research has demonstrated that multimodal signals can reveal meaningful patterns in collaboration, including engagement trajectories, coordination quality, and breakdowns in shared understanding \cite{wohltjen2024interpersonal, yan2024evidence, sharma2021challenging,sharma2020multimodal}. However, most existing work remains focused on retrospective analysis or diagnostic monitoring, rather than using multimodal data to drive \textit{real-time pedagogical adaptation}.
\subsection{{Limitations of Current Adaptive Collaborative Learning Systems}}
Despite growing interest in multimodal and collaborative AIED, most adaptive learning systems remain individual-centric and reactive \cite{sharma2020multimodal}. These systems typically identify suboptimal learner states only after breakdowns have occurred, and often provide alerts or analytics rather than pedagogically grounded adaptive scaffolding \cite{edwards2025human}. As a result, support may arrive too late to prevent misalignment, cognitive overload, or disengagement, leading to inefficient adaptation sequences \cite{moreno2015proactive}.\\
Furthermore, relatively few systems treat collaboration itself as the unit of modelling and instruction. By focusing primarily on individual learner models \cite{kent2020investigating}, existing approaches struggle to support the shared regulatory processes that underpin collaborative learning. This reveals a key gap in AIED, namely the need for systems that can model dyadic learning states, anticipate emerging collaboration breakdowns, and proactively orchestrate instructional support to sustain productive joint activity.
{\subsection{Proactive Adaptivity and Forecast-Driven Instruction in AIED}}
Recent AIED research emphasizes a shift from reactive toward \textit{proactive adaptivity}, in which systems anticipate learners’ future states and intervene before problems fully emerge \cite{mallik2023proactive}. Forecasting models offer a promising mechanism for enabling such anticipatory support, transforming adaptivity from a reactive alert system into a forward-looking instructional decision policy \cite{alkan2025using}.
From an Intelligent Tutoring Systems (ITS) perspective, this enables tighter integration between learner modeling, state prediction, pedagogical decision-making, and adaptive intervention. However, forecast-driven adaptivity has been applied primarily at the individual level \cite{sharma2024self}, leaving open the challenge of using predictive modeling to regulate collaborative learning processes in real time.
\subsection{{Positioning the Present Work-Collaboration as the Object of Instruction}}
To address this gap, we introduce \textbf{ProPACT}, an AI-driven adaptive learning system that treats collaboration itself as the object of instruction. ProPACT constructs a \emph{dyadic learner model} based on Joint Visual Attention (JVA) and Joint Mental Effort (JME), alongside individual mental effort (ME). While ME captures learners’ cognitive load, JVA and JME reflect visual and cognitive alignment, providing direct indicators of collaborative regulation.

Building on this multimodal model, ProPACT uses an XGBoost-based forecasting mechanism to predict suboptimal collaboration states up to 30 seconds ahead. These predictions drive a pedagogically grounded adaptive policy that orchestrates dialogue prompts, gaze-awareness cues, AI-assisted code suggestions, and task-based hints, while fading support when collaboration is productive. This forecast-driven approach advances adaptive learning systems from individual-level monitoring toward real-time regulation of collaborative learning processes.

The present study addresses the following research questions:
\begin{itemize}
    \item \textbf{$RQ_1$:} How does ProPACT affect task performance and code quality in pair programming?
    \item \textbf{$RQ_2$:} How does ProPACT influence dyadic regulation processes?
   
\end{itemize}

Accordingly, this study examines ($RQ_1$) how ProPACT affects task performance and code quality in pair programming, and ($RQ_2$) how it influences dyadic regulation processes, including Joint Visual Attention, Joint Mental Effort, and individual mental effort.

\section{System Architecture and Adaptive Mechanisms}
\subsection{Multimodal Measures (JVA, ME, JME)}

To support real-time classification of collaboration states, all indicators were normalized and discretized into three levels (\emph{High}, \emph{Average}, \emph{Low}) using empirically validated thresholds from prior DUET and eye-tracking research \cite{duchowski2018ipa,schneider2018leveraging,sharma2023cscl}. A two–standard-deviation (2SD) criterion relative to each participant’s resting baseline was applied, with values above $+2\mathrm{SD}$ classified as \emph{High}, below $-2\mathrm{SD}$ as \emph{Low}, and values within $\pm2\mathrm{SD}$ as \emph{Average}. High JVA and JME reflect strong attentional and cognitive alignment, while extreme ME values indicate overload or disengagement.

Joint Visual Attention (JVA) captures the extent to which collaborators attend to the same task-relevant objects over time \cite{sharma2023cscl}. Gaze data were mapped onto a persistent, non-visible grid aligned to the code structure to mitigate distortions from scrolling or resizing. Gaze distributions were aggregated over 30-second windows and compared using cosine similarity to yield a measure of shared attentional focus.

Individual mental effort was measured using the \textit{Index of Pupillary Activity (IPA)} \cite{duchowski2018ipa}, computed from pupil-diameter fluctuations and segmented into non-overlapping 10-second windows. Joint Mental Effort (JME), representing cognitive alignment between dyad members, was computed server-side from the individual mental-effort signals. After discretization into a bounded integer range, synchrony between partners was quantified using cross-recurrence quantification \cite{coco2014cross}, producing a time-synchronized measure of shared cognitive states.

\subsection{Feedback Tools}
\begin{figure*}[t!]
    \centering
    \begin{subfigure}[t]{0.35\textheight}
        \centering
        \includegraphics[width=0.95\linewidth]{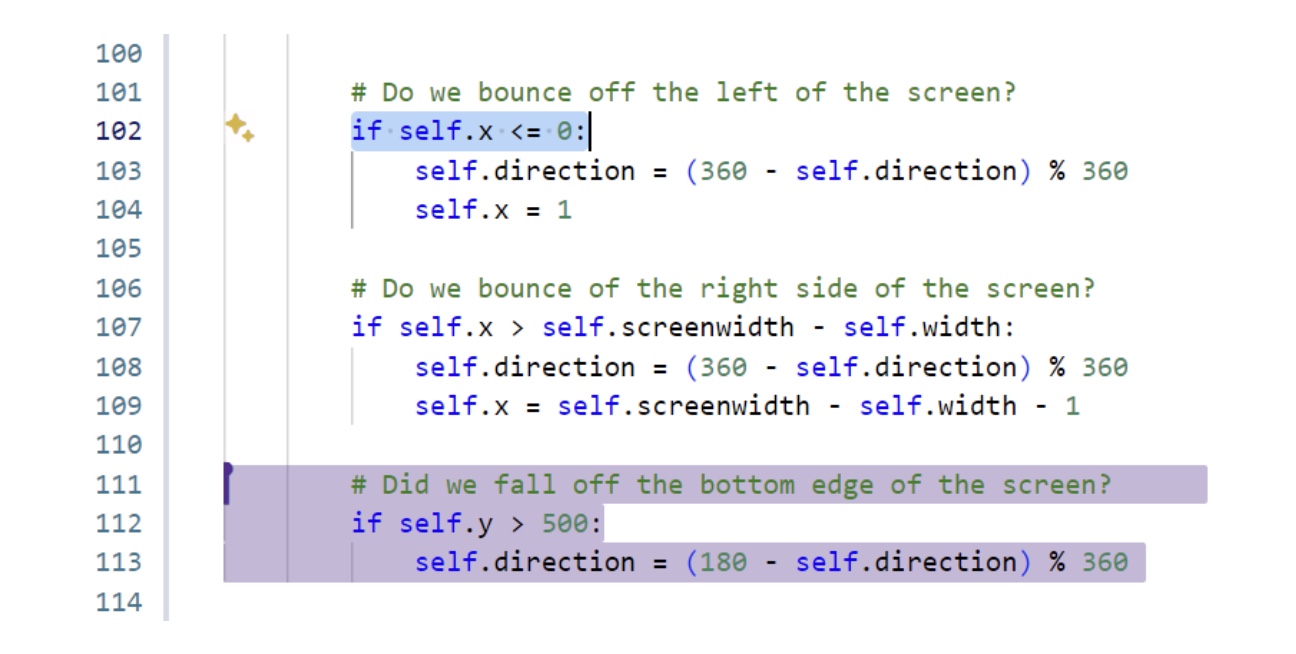}
        \caption{Example of dual text selection: developer A selecting codeline 101 and developer B codeline 111 - 113}
        \label{fig:Dual}
    \end{subfigure}%
    \hfill
    \begin{subfigure}[t]{0.37\textwidth}
        \centering
        \includegraphics[width=\textwidth]{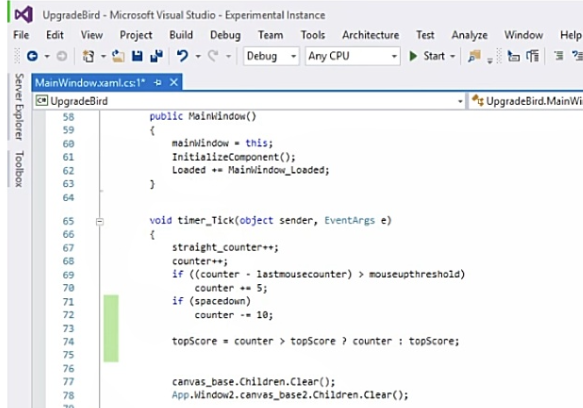}
        \caption{Gaze-awareness tool showing partner’s gaze }
        \label{fig:Gaze}
    \end{subfigure}

   \begin{subfigure}[t]{0.4\textwidth}
        \centering
          \includegraphics[width=\textwidth]{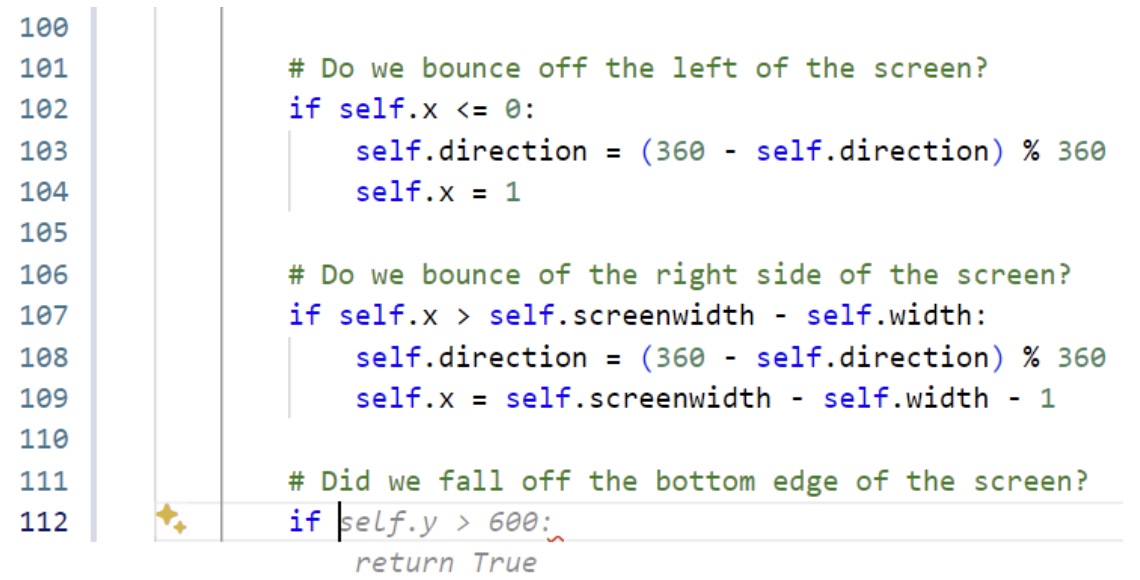}
        \caption{Example of GitHub Copilot’s autocomplete feature providing the correct answer on line 112
 }
         \label{fig:Git}
    \end{subfigure}%
     \hfill
     \begin{subfigure}[t]{0.4\textwidth}
         \centering
         \includegraphics[width=\textwidth]{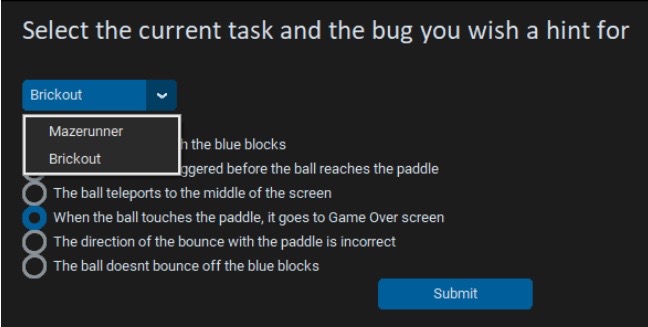}
         \caption{The hint prompt window, user selects the task and bug}
         \label{fig:hint}
    \end{subfigure}

      \begin{subfigure}[t]{0.45\textwidth}
        \centering
         \includegraphics[width=\textwidth]{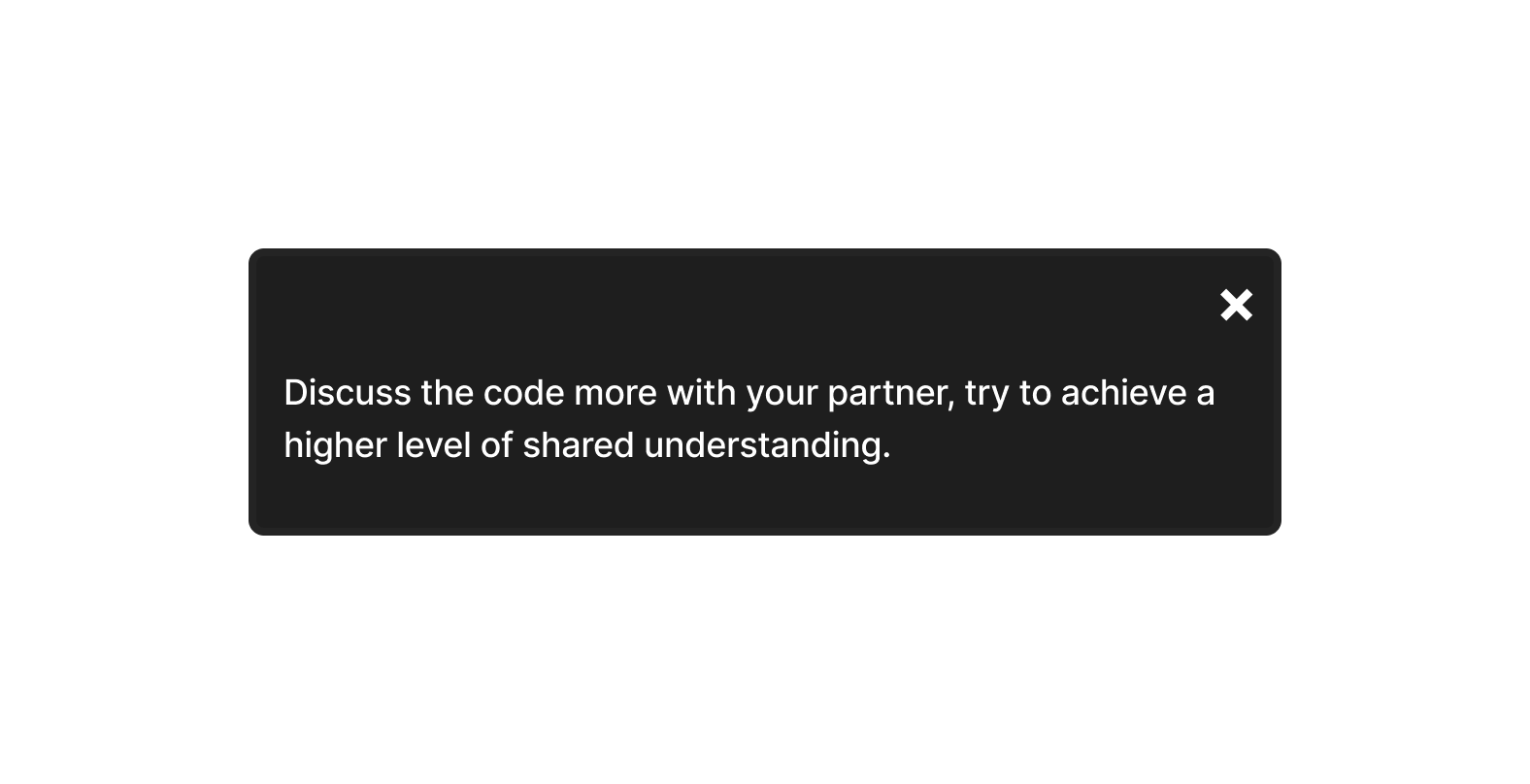}
        \caption{discuss2}
         \label{fig:Disc}
    \end{subfigure}%
     \hfill
    
    \caption{Feedback Tools}
    \label{fig:FTools}
\end{figure*}

Figure \ref{fig:FTools} demonstrates the five feedback tools integrated in the system.

\label{Sec:Tools}
\begin{itemize}
\item \textbf{GitHub Copilot:} Figure \ref{fig:Git} illustrates this feedback. In our study, GitHub Copilot is restricted to its AI-driven autocomplete functionality, providing context-sensitive code suggestions within the editor. Prior work shows that Copilot can speed task completion \cite{shihab2025effects} and reduce perceived cognitive effort \cite{choi2025improving}, though its learning effects remain unclear. In our framework, enabling Copilot refers to temporary activation of autocomplete as a scaffold when forecasted mental-effort patterns indicate emerging cognitive strain.
\item\textbf{Dual Text Selection:} 
Broadcasted text selection continuously shares each collaborator’s cursor selection, providing a lightweight cue for rapid joint focus \cite{jermann2012effects}. Supported natively in collaborative editors such as Visual Studio Code Live Share, this always-on mechanism maintains mutual awareness and aligns with Joint Visual Attention (JVA) as a key indicator of coordination in synchronous collaboration (Figure \ref{fig:Dual}).
\item \textbf{ Gaze-Awareness Tool:}  
Gaze visualization supports remote pair programming by making a partner’s focus visible and improving communication efficiency \cite{d2021shared,d2017improving,hayashi2020gaze}. Building on this work, we visualize a partner’s gaze as a translucent colored rectangle highlighting the current attention region ( nine lines of code), triggered when Joint Visual Attention (JVA) is low (Figure \ref{fig:Gaze}).
\item \textbf{Dialogue Prompt:} 
Low behavioral or cognitive synchrony often signals misaligned understanding or uneven effort in collaboration \cite{sharma2023happens}. Drawing on dual eye-tracking research on mutual awareness, timely conversational prompts can restore alignment when attention or cognitive load diverges \cite{sharma2021challenging}. Accordingly, feedback is triggered when Joint Mental Effort (JME) falls below a threshold, and a small, unobtrusive prompt appears in the editor (Figure~\ref{fig:Disc}) to encourage brief dialogue and re-align mental effort with minimal disruption.
\item \textbf{ Task-Based Hint :} The task-based hint is the most directive form of support and is used only as a last-resort intervention, triggered when less intrusive scaffolds fail and both collaborators exhibit sustained extreme mental-effort levels, indicating cognitive strain or stagnation. This staged design follows adaptive support principles that recommend escalating assistance only after lower-intensity interventions are exhausted, particularly under high cognitive load \cite{marwan2020unproductive,xie2017more}. When activated, the system prompts selection of the current task and bug and then provides a targeted hint (Figure \ref{fig:hint}) that directs attention to relevant code regions and clarifies the underlying issue to restore cognitive balance.

    \end{itemize}

\subsection{Forecasting and Feedback Trigger Logic}

\begin{figure}
    \centering
    \includegraphics[width=0.99\linewidth]{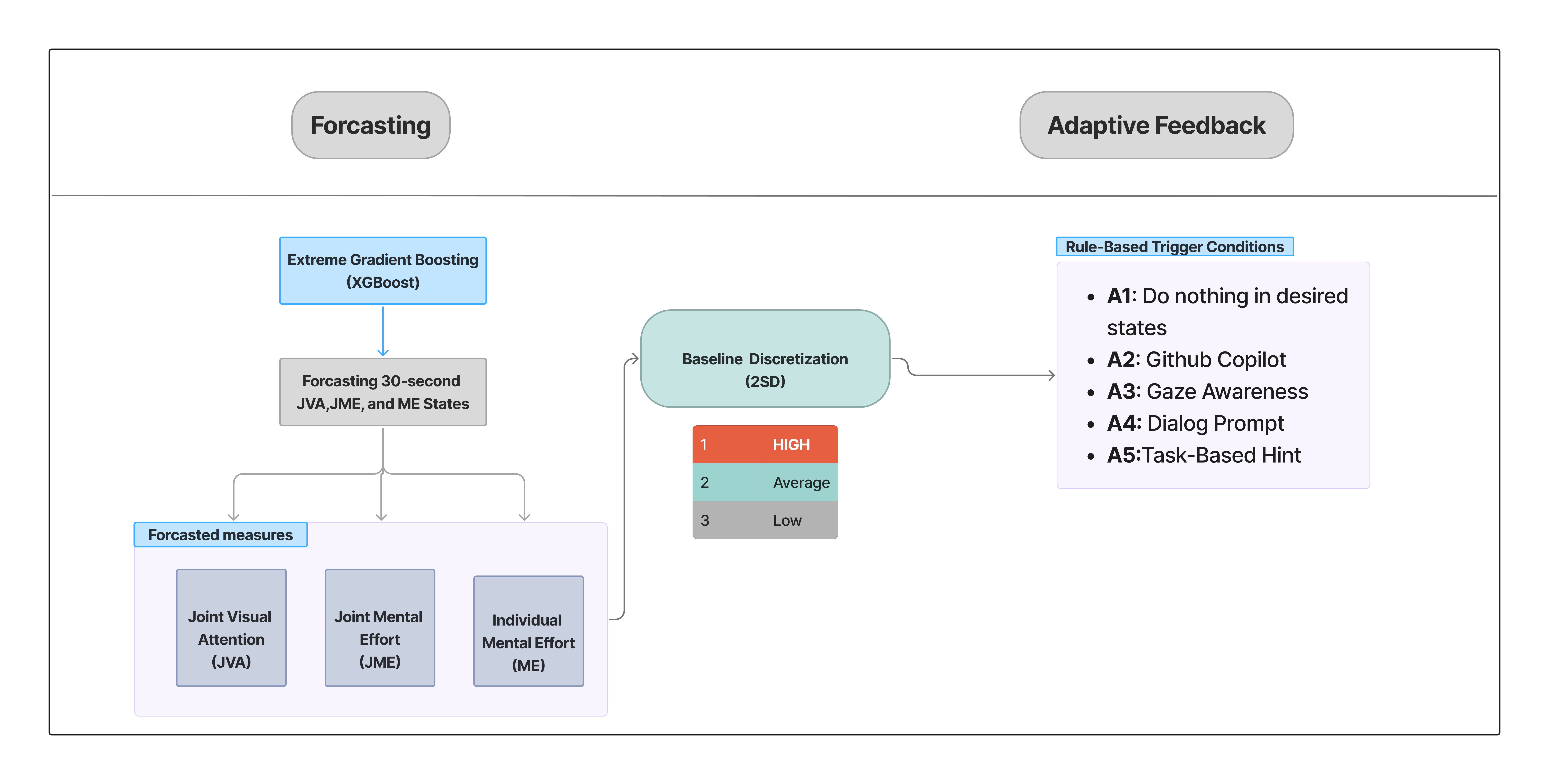}
    \caption{Hybrid AI Framework.}
    \label{fig:framework}
\end{figure}

The feedback system is implemented as a Hybrid-AI framework (Figure~\ref{fig:framework}) that combines data-driven forecasting with rule-based pedagogical decision-making. An XGBoost model predicts Joint Visual Attention (JVA), Joint Mental Effort (JME), and individual Mental Effort (ME) over a 30-second horizon. Predictions are discretized relative to each participant’s resting baseline using a ±2SD criterion and categorized as \emph{High}, \emph{Average}, or \emph{Low}. These states are compared against a desired collaboration matrix and mapped to predefined trigger conditions, with feedback selected via a top-down hierarchical policy that prioritizes minimal intervention and escalates support only when forecasted risk of breakdown increases. Trigger conditions are summarized in Table~\ref{tab:trigger}.

\begin{table}[t]
\footnotesize
\centering
\caption{Adaptive feedback types and their trigger conditions}
\label{tab:feedback_triggers}
\begin{tabular}{p{6cm} p{7cm}}
\hline
\textbf{Feedback type} & \textbf{Trigger condition} \\
\hline
A1: Do nothing (Desired state) &
 ($ME_s$ = AVG)  \& (JVA = H) \& (JME = H) \\

A2: GitHub Copilot  &
 $ME_s$ = HH or LL, or ($ME_s$ = HL and JVA = L) \\

A3: Gaze-awareness &
JVA = Low \\

A4: Dialog prompt &
JME = Low \\

A5: Task-based hint &
Both $ME_s$ = High \\
\hline
\label{tab:trigger}
\end{tabular}
\end{table}

\subsection{Empirical Evaluation}

\subsubsection{Feedback Effectiveness Measures:}

\begin{enumerate}
    \item \textbf{Debugging success:} This metric measures the number of bugs successfully fixed within the allotted time, including the bug actively being addressed at task completion.
    \item \textbf{Debugging time on task:} The total time on task is the total time spent in seconds. This metric measures the efficiency of the collaboration.
    \item \textbf{Feedback uptake:} This measure is defined, using code snippets, as the changes made to the code by the pair after receiving a given piece of feedback and before the next feedback is triggered. It operationalizes learner agency  by capturing the deliberate effort programmers invest in modifying the shared artifact during collaboration. 
\end{enumerate}

\subsubsection{Participants and Procedure:}
We used a within-subjects design with two conditions: no feedback (control) and \textit{ProPACT} feedback. Twenty-six pairs (19 female, 33 male) of undergraduate or master’s students in computer science or engineering at a European university participated, all having completed an introductory programming course.

Pairs were randomly assigned and provided informed consent. Participants received a brief system demonstration, after which eye trackers were individually calibrated. They were informed that the programming tasks contained only logical bugs and no syntax errors. 
\subsubsection{Data Analysis:}

To address $RQ_1$, we conducted a paired t-test with the debugging success metrics as the dependent variable and the feedback conditions as the independent variable. To address RQ2, we conducted a paired t-test with the JVA and JME, before and after providing the feedback (JVA and JME were aggregated two minutes prior and two minutes after the feedback was provided). Prior to conducting t-tests, we verified the homoscedasticity of the variables using Breusch-pegan Test; and the normality using Shapiro-Wilk test. 
\section{Results}
First, we analysed the order-effect on all the three measurements for feedback efficiency because this experiment was a within-subject design, and we had balanced the order of the control task and the experimental task. We observed no order effect on any of the dependent variables i.e., debugging success (t[24] = 0.45, p $>$ .05), debugging time on task (t[24] = 0.56, p $>$ .05), and the feedback uptake  (t[24] = 1.01, p $>$ .05).

\begin{figure*}[t!]
    \centering
    \begin{subfigure}[t]{0.5\textwidth}
   \centering
    \includegraphics[width=0.9\linewidth]{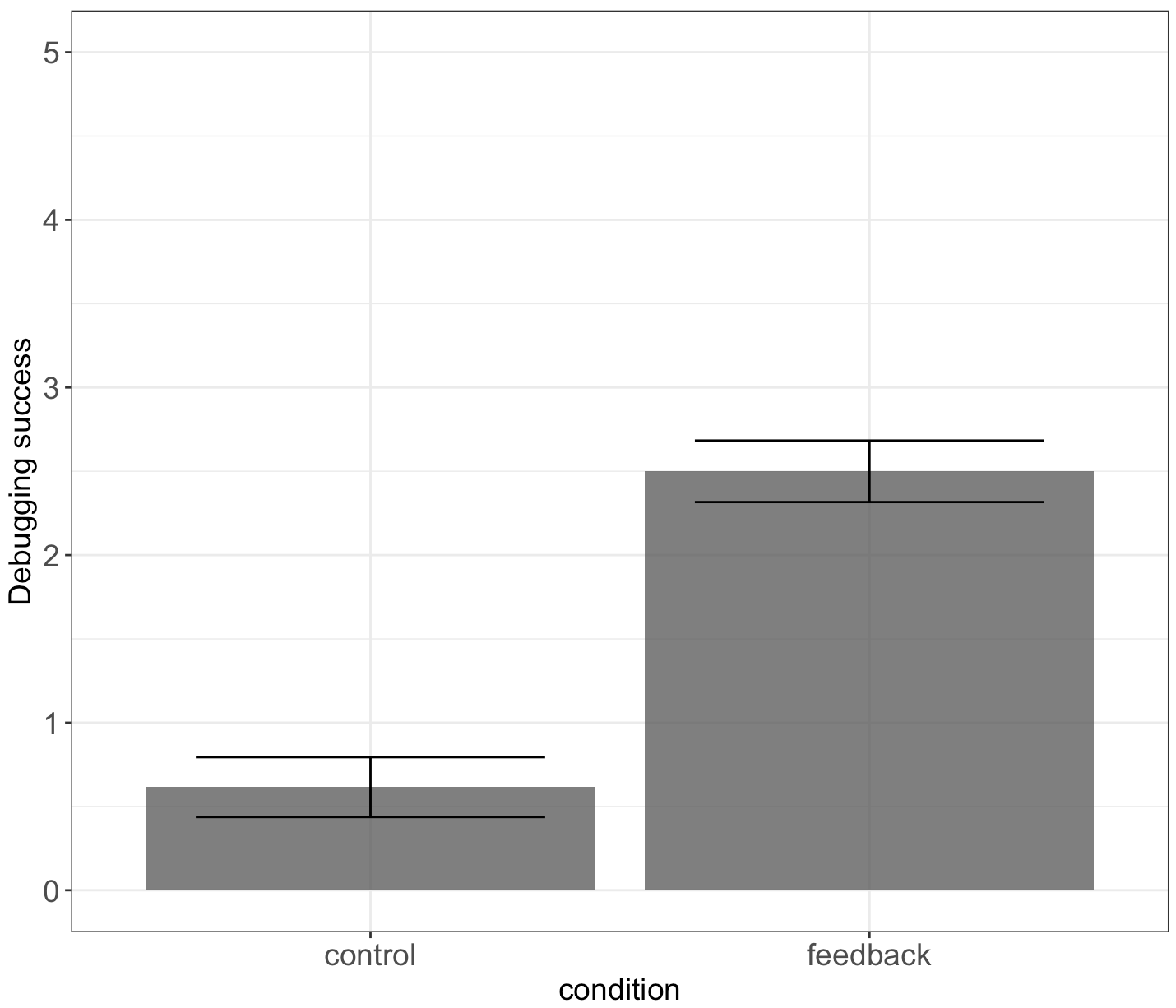}
    \caption{Debugging success across the proactive feedback conditions.}
    \label{fig:proactive-success}
    \end{subfigure}%
    \hfill
    \begin{subfigure}[t]{0.5\textwidth}
        \centering
    
     \includegraphics[width=0.9\linewidth]{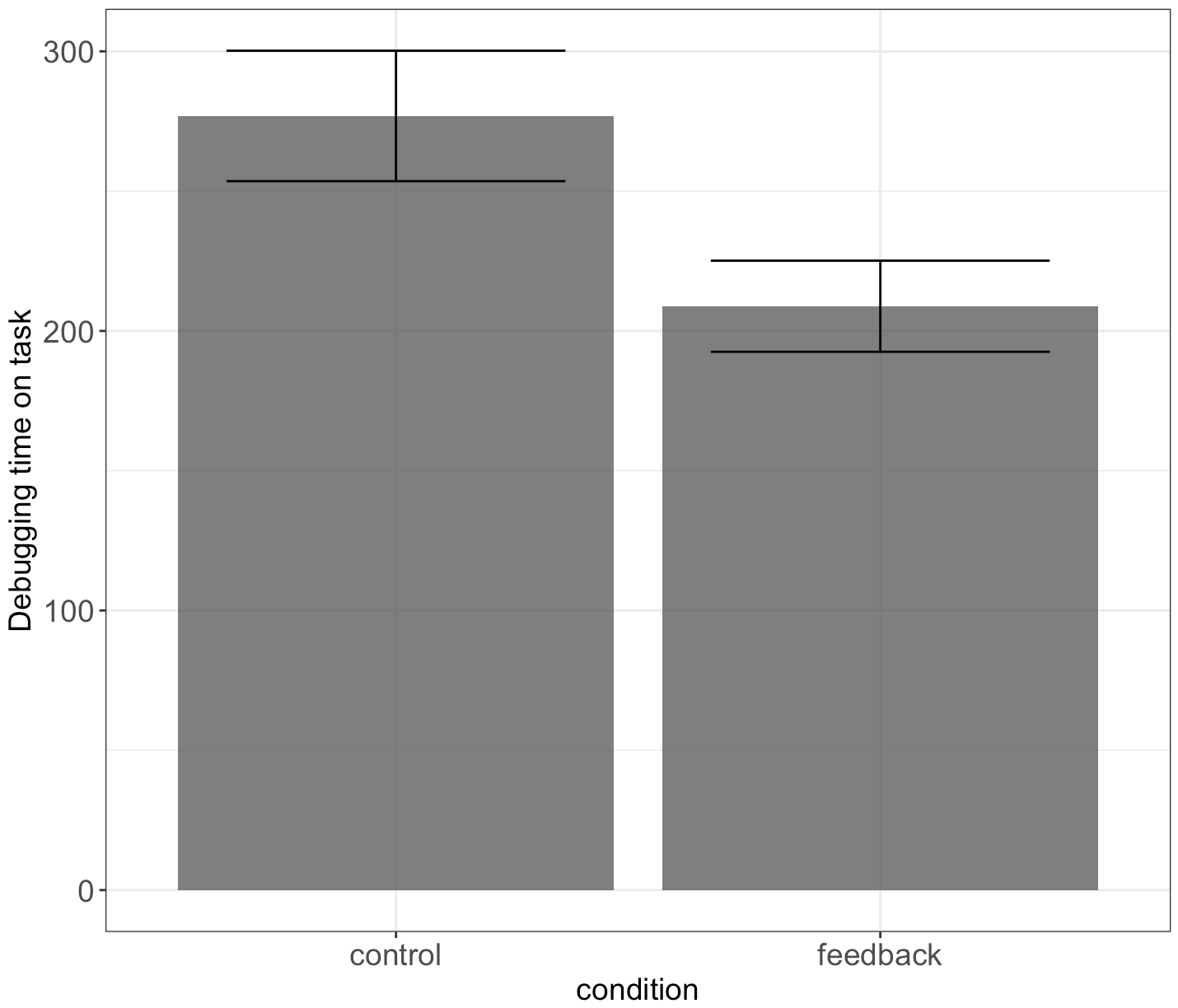}
    
    \caption{Debugging time on task across the proactive feedback conditions.}
    \label{fig:proactive-tot}
    \end{subfigure}
    \begin{subfigure}[t]{0.99\textwidth}
         \includegraphics[width=0.4\linewidth]{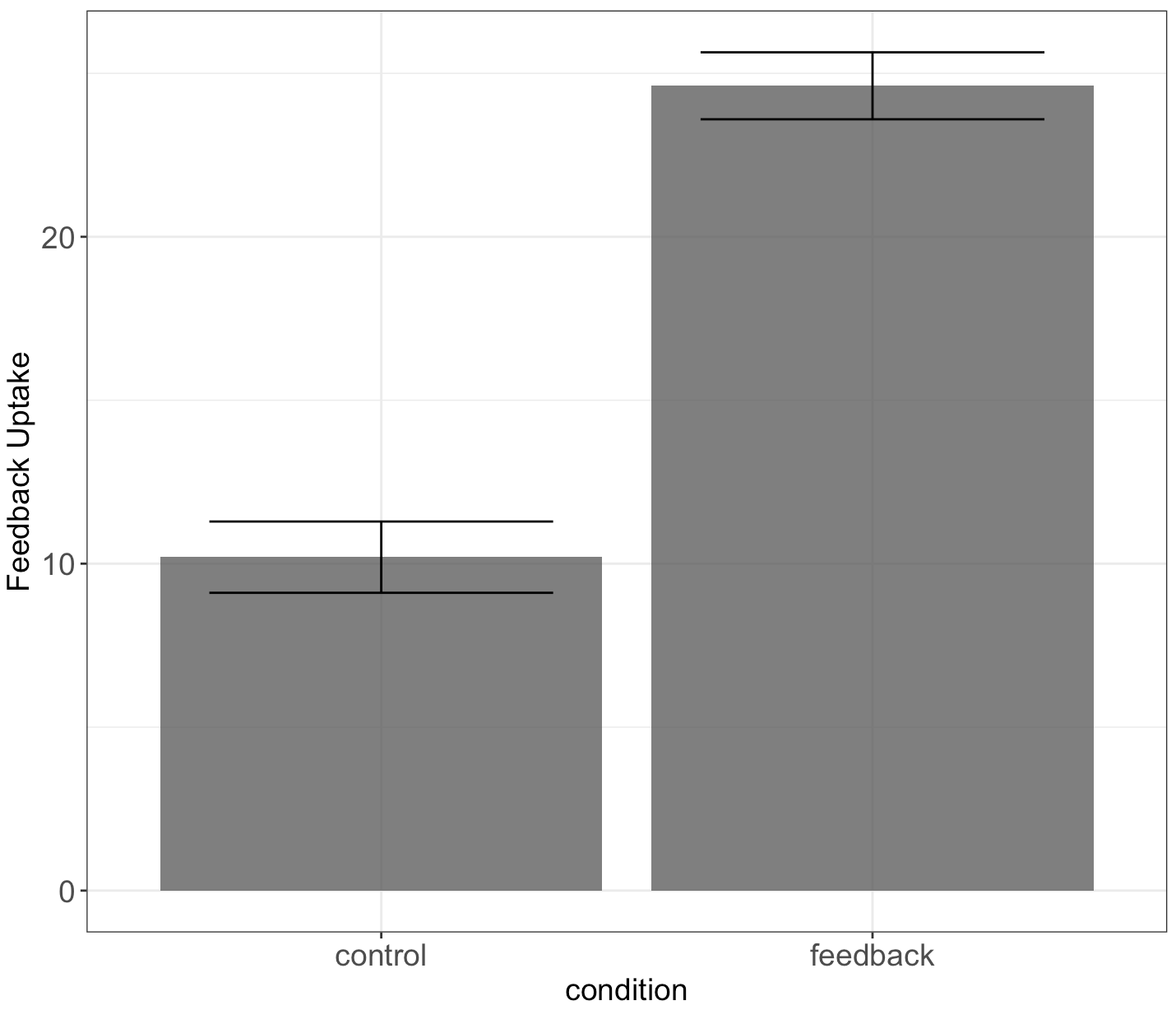}
    \caption{Feedback uptake across the proactive feedback conditions.}
    \label{fig:proactive-uptake}
    \end{subfigure}
\hfill
    \label{FTools}
    \caption{Debugging \textit{success} , \textit{time}, and \textit{uptake} across the proactive feedback conditions ($RQ_1$)}
\end{figure*}

To address $RQ_1$ (feedback condition vs pair-programming performance), a t-test with the feedback category as the independent variable and debugging success as the dependent variable shows a clear difference in the debugging success among the feedback categories (t[49.96] = -13.51, p $<$ .0001, figure \ref{fig:proactive-success}). The debugging success is significantly higher in the feedback condition than in the control condition. Another t-test with the feedback category as the independent variable and debugging time on task as the dependent variable shows a clear difference in the debugging time on task among the feedback categories (t[44.70] = 4.39, p $<$ .0001, figure \ref{fig:proactive-tot}). The debugging time on task is significantly lower in the feedback condition than the control condition.

\begin{figure*}[t!]
    \centering
    \begin{subfigure}[t]{0.49\textwidth}
       \centering
    \includegraphics[width=0.9\linewidth]{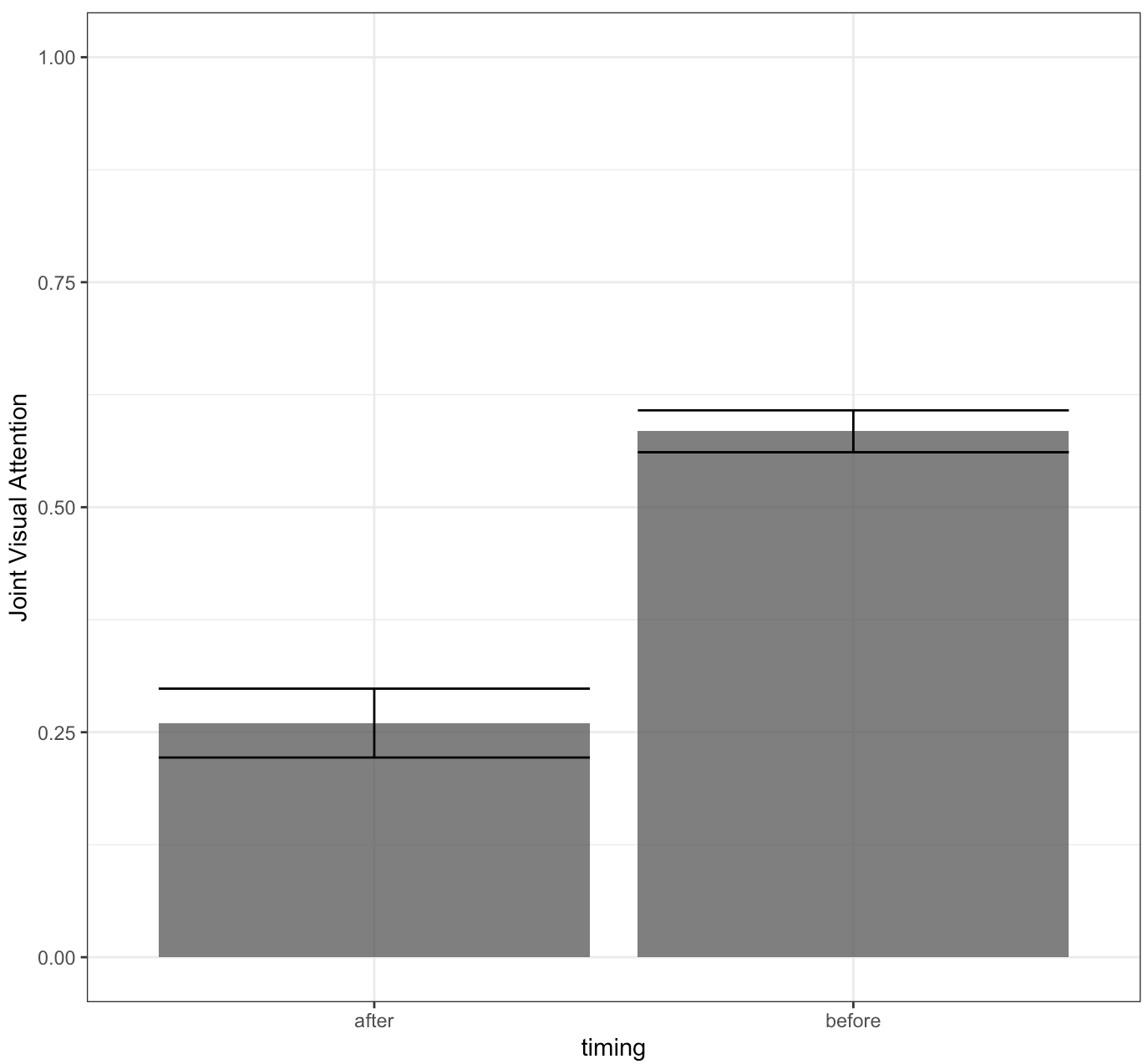}
    \caption{JVA values before and after providing the feedback.}
    \label{fig:JVA-before-after}
    \end{subfigure}%
    \hfill
    \begin{subfigure}[t]{0.49\textwidth}
       \centering
    \includegraphics[width=0.9\linewidth]{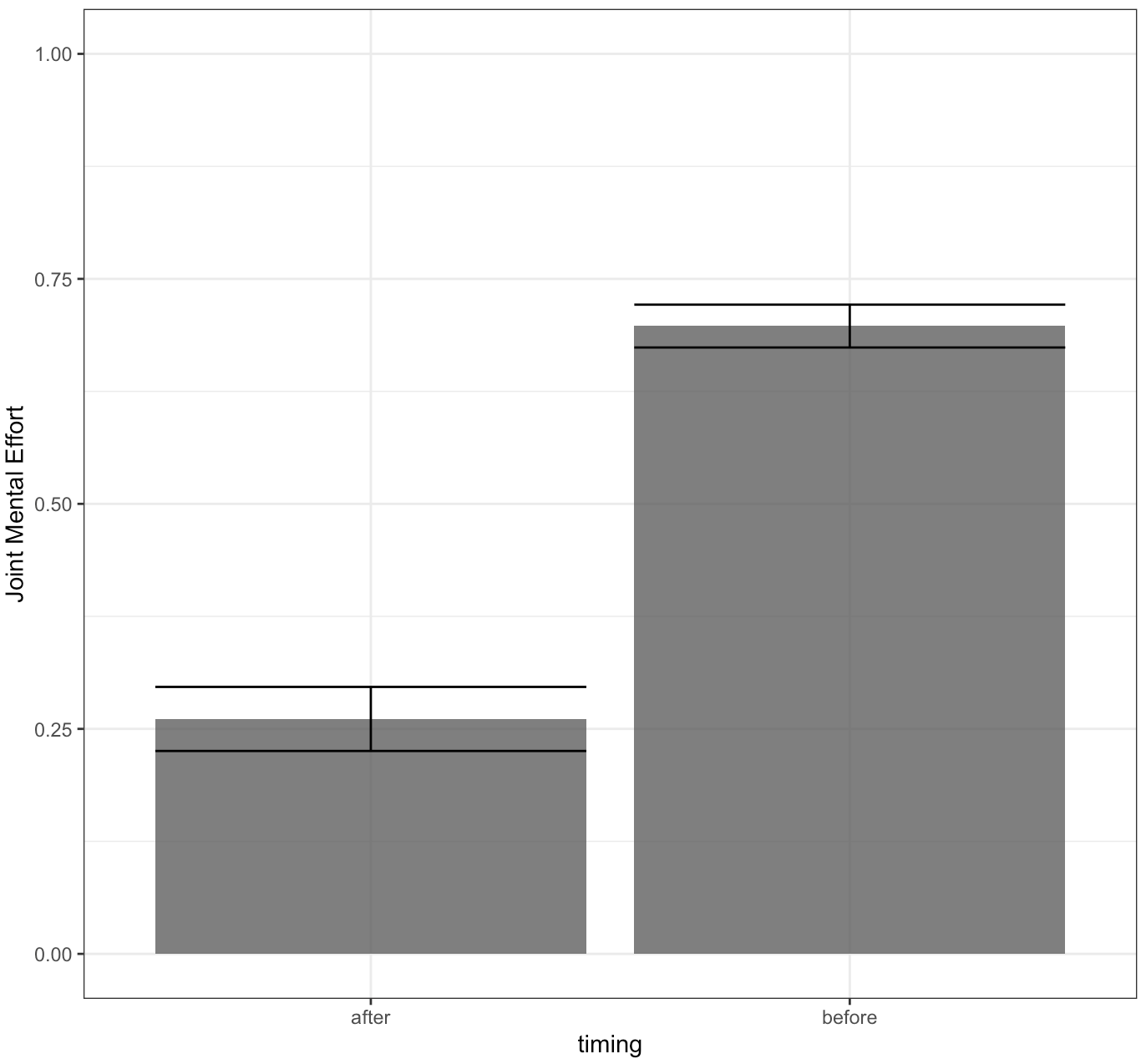}
    \caption{JME values before and after providing the feedback.}
    \label{fig:JME-before-after}
    \end{subfigure}

    \hfill
    \label{FTools}
    \caption{Process Measurements (RQ2)}
\end{figure*}

Finally, the last t-test with the 'feedback category' as the independent variable and 'feedback uptake' as the dependent variable shows a clear difference in the feedback uptake among the feedback categories (F[49.81] = -17.69, p $<$ .0001, figure \ref{fig:proactive-uptake}). The feedback uptake on task is significantly higher in the feedback condition than the control condition.
To address $RQ_2$ (before and after feedback vs process measures), a test between the JVA values before (average for the two minutes before) and after (averaged for two minutes after) the feedback was provided showed a significant increase in the JVA of the dyads (t[df = 25] = 12.76, p $<$ .0001, figure \ref{fig:JVA-before-after}). Similarly a test between the JME values before (average for the two minutes before) and after (averaged for two minutes after) the feedback was provided showed a significant increase in the JME of the dyads (t[df = 25] = 19.33, p $<$ .0001, figure \ref{fig:JME-before-after}).

\section{Discussion}
{Building on previous systematic analysis of regulation in collaborative learning systems \cite{sharma2024self}, which highlights the predominance of individual-level metrics over group and multi-level regulation,} this study investigated the effectiveness of ProPACT, the AI-driven, proactive adaptive learning system  that models dyadic learner states \textit{performance} and \textit{collaborative processes} in pair programming learning activities.
By treating collaboration itself as the object of instruction, rather than focusing solely on individual learners or task performance, and by intervening based on forecasted rather than observed breakdowns, our system advances adaptive support from individual-level monitoring toward \textit{dyadic regulation of collaborative learning}. \\\newline
\textbf{Effects on Pair-Programming Performance: }
Addressing $RQ_1$, the results show that proactive multimodal feedback significantly improved pair-programming performance: pairs supported by the adaptive system resolved more bugs and completed debugging tasks more efficiently than those in the control condition.
{These findings empirically support claims that learning regulation can be triggered by early cognitive or behavioral signals rather than only after overt breakdowns occur \cite{jarvela2023human}. While prior work has focused largely on design principles and exploratory prototypes \cite{edwards2024mai,nguyen2022exploring}, evidence that such early signals yield measurable performance gains in authentic collaborative tasks has remained limited.}
{Research comparing reactive and proactive feedback designs highlights the critical role of intervention timing, with anticipatory support offering advantages for sustaining productive collaboration \cite{kim2025socially,edwards2024mai}. Consistent with this perspective, our findings indicate that anticipatory regulation sustains debugging momentum while preserving autonomy, enhancing both task efficiency and interaction quality through improved alignment of attention, effort, and strategy.}
{A learner-orchestrated feedback perspective provides a useful lens for interpreting these results \cite{wood2025empowering}. The observed increase in feedback uptake suggests that the interventions supported regulatory decisions without undermining autonomy, aligning with calls for AI-supported feedback that preserves agency by integrating seamlessly into ongoing learning activity \cite{li2025ai,rakhmetov2025evaluation}.}
Importantly, performance gains were accompanied by increased feedback uptake, indicating active engagement with the scaffolds rather than passive compliance. Together, these results support framing proactive adaptive feedback as an instructional decision policy that moves beyond reactive alerts toward pedagogically grounded orchestration of collaborative learning.\\\newline
\textbf{Effects on Collaborative Processes: }
Addressing $RQ_2$, the findings show significant post-intervention increases in both JVA and JME, providing process-level evidence that the system improved collaboration quality rather than merely accelerating task completion. Higher JVA reflects stronger coordination of attention toward shared task-relevant information, while higher JME indicates improved alignment of cognitive engagement between partners. These results align with theories of socially shared regulation emphasizing shared understanding and cognitive balance \cite{edwards2025human}, and suggest that proactive scaffolds support dyadic re-alignment before coordination breakdowns escalate. 
\subsection{Limitations and Future Work}
Several limitations should be acknowledged.
\textit{First}, the study was conducted in a controlled laboratory setting with a relatively small sample size and short-duration debugging tasks. While this enabled precise measurement of multimodal collaboration signals, it may limit generalizability to classroom, remote, or industrial software development contexts, where collaboration unfolds over longer timescales and under less structured conditions.
\textit{Second}, the system relies on specialized sensing infrastructure, including dual eye-tracking and pupillometry. Although these sensors provide fine-grained insight into attention and cognitive effort, they pose practical and financial constraints for large-scale deployment. Future work should explore whether comparable adaptive models can be supported using more scalable signals, such as interaction logs, keystroke dynamics, webcam-based gaze estimation, or wearable and ambient sensing technologies.

Beyond technical scalability, future research should examine longer-term learning outcomes, including conceptual understanding, retention, and transfer beyond immediate debugging performance.

Longitudinal studies could clarify whether repeated exposure to proactive collaborative scaffolding supports durable self-regulation and metacognitive awareness.

Further work is also needed to evaluate the system across diverse tasks, domains, expertise levels, and collaboration structures. Extending beyond pair programming to small-group or team-based settings would enable investigation of how multimodal forecasting and adaptive feedback scale to more complex group dynamics. Finally, future studies should explore how learners interpret, trust, and appropriate proactive feedback over time, including how perceptions of autonomy, transparency, and fairness evolve—key considerations for designing human-centered, pedagogically grounded AIED systems.

\bibliography{sn-bibliography}

\end{document}